\documentclass{article}
\usepackage{smc}
\usepackage{times}
\usepackage{ifpdf}
\usepackage[english]{babel}
\usepackage{cite}

\usepackage[nolist]{acronym}
\usepackage{amsmath}
\usepackage{xcolor}

\begin{acronym}
    \acro{mir}[MIR]{Music Information Retrieval}
    \acro{mfcc}[MFCC]{Mel Frequency Cepstrum Coefficient}
    \acro{stft}[STFT]{Short-Time Fourier-Transform}
    \acro{kl}[KL]{Kullback-Leibler}
    \acro{cw}[C\&W]{Carlini \& Wagner}
    \acro{snr}[SNR]{Signal-to-Noise ratio}
    \acro{mp}[MP]{Mutual Proximity}
\end{acronym}

\DeclareMathOperator{\sign}{sign}
\DeclareMathOperator{\clip}{clip}
\DeclareMathOperator{\bt}{bt}

\def\papertitle{Defending a Music Recommender Against Hubness-Based Adversarial Attacks}
\def\firstauthor{Katharina Hoedt}
\def\secondauthor{Arthur Flexer}
\def\thirdauthor{Gerhard Widmer}

\newif\ifpdf
\ifx\pdfoutput\relax
\else
   \ifcase\pdfoutput
      \pdffalse
   \else
      \pdftrue
\fi

\ifpdf 
  \usepackage[pdftex,
    pdftitle={\papertitle},
    pdfauthor={\firstauthor, \secondauthor, \thirdauthor},
    bookmarksnumbered, 
    pdfstartview=XYZ 
   ]{hyperref}

  \usepackage[pdftex]{graphicx}
  \graphicspath{{./figures/}}
  \DeclareGraphicsExtensions{.pdf,.jpeg,.png}

  \usepackage[figure,table]{hypcap}

\else 
  \usepackage[dvips,
    bookmarksnumbered, 
    pdfstartview=XYZ 
  ]{hyperref}  

  \usepackage[dvips]{epsfig,graphicx}
  \graphicspath{{./figures/}}
  \DeclareGraphicsExtensions{.eps}

  \usepackage[figure,table]{hypcap}
\fi

\hypersetup{
    colorlinks,%
    citecolor=black,%
    filecolor=black,%
    linkcolor=black,%
    urlcolor=black
}

\title{\papertitle}

\oneauthor
   {\firstauthor$^1$ \ \secondauthor$^1$ \ \thirdauthor$^{1, 2}$} 
   {$^1$ Johannes Kepler University Linz, Austria \\ 
   $^2$ LIT AI Lab, Linz Institute of Technology, Austria \\ %
     {\tt \href{mailto:katharina.hoedt@jku.at}{katharina.hoedt@jku.at} \href{mailto:arthur.flexer@jku.at}{arthur.flexer@jku.at}}}

\begin{document}
\capstartfalse
\maketitle
\capstarttrue

\begin{abstract}
Adversarial attacks can drastically degrade performance of recommenders and other machine learning systems, resulting in an increased demand for defence mechanisms. We present a new line of defence against attacks which exploit a vulnerability of recommenders that operate in high dimensional data spaces (the so-called \emph{hubness problem}). We use a global data scaling method, namely Mutual Proximity (MP), to defend a real-world music recommender which previously was susceptible to attacks that inflated the number of times a particular song was recommended. We find that using MP as a defence greatly increases robustness of the recommender against a range of attacks, with success rates of attacks around $44\%$ (before defence) dropping to less than $6\%$ (after defence). Additionally, adversarial examples still able to fool the defended system do so at the price of noticeably lower audio quality as shown by a decreased average SNR.
\end{abstract}
%


\section{Introduction}
\label{sec:Introduction}

Adversarial examples were previously reported in various fields of application (cf. \cite{Szegedy2014IntruigingProperties,Carlini2018STT,Sturm2014DetermineHorse}) as small perturbations of input data that drastically change the performance of machine learning systems. Since then, numerous attempts to make systems more robust and \emph{defend} them against these attacks have been made (cf. \cite{Xu2018DetectingAdversaries,Jakubovitz2018DefenceJacobianRegularization,Madry2018AdversarialTraining}). 

In previous work~\cite{Tismir2021}, a successful attack on the music recommender FM4 soundpark~\cite{Gasser2009Soundpark} was presented, inflating the number of times (perturbed) songs were recommended by the system. The paper failed, however, to provide an outlook on how this attack could be weakened and did not present a method to defend against the attack. The proposed attack exploited \emph{hubness}, a problem of learning in high dimensions that leads to some songs being recommended very often and other songs being never recommended~\cite{Flexer2018MutualProximity}. As the attack amplifies the negative effect of hubness on the recommendation of songs, finding a defence against this kind of attacks would contribute positively towards the fairness of a recommender system.

To explain the existence of hub points, it was previously shown that for any high but still finite dimensionality, some points are expected to be closer to the center of all
data (or local center in case of multimodal data) than other points and are at the same time closer, on average, to all other points~\cite{Radovanovic2010Hubs}. Such hub points appear in nearest neighbour lists of many other points, resulting in asymmetric neighbour relations: a hub $y$ is the nearest neighbour of $x$, but the nearest neighbour of the hub $y$ is another point $a$ $(a \neq x)$. One approach to repair asymmetric neighbourhood relations and hence mitigate the hubness problem is \ac*{mp}~\cite{Schnitzer2012MutualProximity}, which globally scales distances (i.e., considers neighbourhood information of \emph{all} objects), and transforms the distance between two objects into a measure that captures how similar the neighbourhoods of these two objects are. 

In this work, we aim at finding a defence against adversarial attacks that exploit the hubness issue present in various recommender systems. Our contribution in this paper is threefold: (i) for the first time we utilise the hubness-reduction method \acs*{mp} as a defence against adversarial attacks; 
(ii) in order to create a more difficult defence scenario for \acs*{mp}, we additionally incorporate
knowledge of the defence into a modified adversarial attack; and (iii) we also investigate \acs*{mp} as a post-hoc defence, i.e., using \acs*{mp} to post-process recommendations issued by an attacked but undefended system.


\section{Related Work}
\label{sec:Related Work}
Hubness was first described in \ac*{mir}~\cite{Pachet2004FirstHubness}, but is now acknowledged as another aspect of the curse of dimensionality, and hence a general machine learning problem~\cite{Radovanovic2010Hubs}. It was not only demonstrated to be a relevant problem in audio-based recommender systems~\cite{Gasser2009Soundpark,Flexer2018MutualProximity} (which are our focus here), but also in recommenders based on collaborative filtering~\cite{Knees2014HubnessCollaborativeFiltering,hara2015reducing} as well as in general multimedia retrieval~\cite{Schnitzer2014ECIR}.

To reduce the effect of the hubness phenomenon, various hubness mitigation techniques were previously proposed. Feldbauer and Flexer differentiate between methods that use \emph{dimensionality reduction}, centring-based methods that \emph{reduce spatial centrality}, methods that aim at \emph{repairing asymmetric relations} (e.g., \acs*{mp}), or using entirely \emph{different distance measures}~\cite{Feldbauer2019HubnessReductionComp}. We look at \acs*{mp} in particular in this work, as it was previously applied to data of the FM4 soundpark~\cite{Flexer2018MutualProximity}, has been shown to be very effective at reducing hubness in diverse datasets \cite{Feldbauer2019HubnessReductionComp} and is applicable even for large amounts of data with efficient approximations~\cite{Feldbauer2018FastMP}.

Defences against adversarial attacks can broadly be divided into approaches in which adversaries are either \emph{reduced} or \emph{detected} (cf. \cite{Xu2018DetectingAdversaries}) in order to make a system more robust~\cite{Huang2020defencesurvey}. One of the most notable approaches is adversarial training~\cite{Madry2018AdversarialTraining}, where adversarial examples are included in the training procedure of a system. As this is a complex task due to the necessity of computing adversarial examples during training time, multiple variants of this defence were proposed more recently (e.g., \cite{Shafahi2019FreeAdversarialTraining,Zhang2019YOPO}) with a focus on efficiency.

Adversarial defence methods are generally viewed to be at an arms race with adversarial attacks, as a lot of defences are only considered purposeful until a new attack is proposed which the method fails to defend against (cf.~\cite{Huang2020defencesurvey}). In this work, we therefore try to apply a method for defending against adversaries that does not require full knowledge of the specific attack; the one limitation we work with, however, is that the attack needs to exploit the hubness phenomenon. Note also that the real-world recommender that was attacked in previous work~\cite{Tismir2021} would not permit adversarial training as a defence method due to the nature of the system (cf. section~\ref{subsec:The Music Recommender System}), as no training in the sense of learning model-parameters is performed. 

In this work, we build upon two approaches that were previously published; first, we use the attack scenario proposed in~\cite{Tismir2021} in order to provide the setting for our defence. Secondly, we apply the hubness-reduction method introduced in~\cite{Flexer2018MutualProximity} and investigate its suitability as a defence method against adversarial examples. This new application is investigated in this work for the first time; in addition to that, we introduce an adapted attack which incorporates knowledge of \acs*{mp} and is hence equipped with more knowledge about the defence to test the boundaries of the proposed defence method.


\section{Data}
\label{sec:Data}
The data we use in this work consists of 5,000 songs from the FM4 Soundpark\footnote{\url{https://fm4.orf.at/soundpark}}, a music discovery site that provides a platform where (lesser known) artists can upload and present their music free of charge. Website visitors can listen to and download all the music at no cost. Songs are generally assigned to one or two out of six different genres~\cite{Gasser2009Soundpark}. The songs have a duration of at least $2$ and a maximum of $22.5$ minutes, with an average of $4.1$ minutes.


\section{Methods}
\label{sec:Methods}
\subsection{The Music Recommender}
\label{subsec:The Music Recommender System}
The system we adapt in this work is the audio-based recommender system integrated in the FM4 Soundpark \footnote{Currently not accessible due to Adobe Flash Player's phaseout beginning of 2021.}~\cite{Gasser2009Soundpark}, which was previously attacked in related work~\cite{Tismir2021}. To prepare the data for the recommender system, every audio file is converted to mono and re-sampled to 22.05 kHz. The central two minutes of a song are further transformed to \acp*{mfcc} by applying a \acl*{stft} (Hann window with a window size of 1,024 and a hop size of 512), a transformation to Mel-scale and finally using a discrete cosine transform to compress the results to 20 \acsp*{mfcc}.

After the pre-processing step every song is represented by a multivariate Gaussian, which is estimated on the respective  \acsp*{mfcc} \cite{Gasser2009Soundpark}. To approximate the similarity of two songs in the next step, the Gaussians describing them are used to compute a symmetrised \ac*{kl} divergence (cf. \cite{Gasser2009Soundpark}), i.e., 

\begin{equation}
    D_{\mathrm{SKL}}(\mathcal{G}_{x}, \mathcal{G}_t) = \frac{\mathrm{KL}(\mathcal{G}_{x} || \mathcal{G}_{t}) + \mathrm{KL}(\mathcal{G}_{t} || \mathcal{G}_{x})}{2}.
    \label{eq:kl}
\end{equation}

Here $\mathcal{G}_{x}$ denotes the Gaussian computed to represent a song $x$ and $\mathrm{KL}$ is the \acs*{kl} divergence of two Gaussian distributions. The symmetrised \acs*{kl} divergence allows us to determine the $k$ nearest neighbours of any song using $D_{\mathrm{SKL}}$ as a distance measure. A recommendation for a particular song consists of these $k$ nearest neighbours, where $k$ is set to $5$ as in the real-world recommender~\cite{Gasser2009Soundpark}.

Note that for this recommender, adding a new song $x'$ requires the computation of its Gaussian $\mathcal{G}_{x'}$, as well as obtaining the symmetric \acs*{kl} divergence between $\mathcal{G}_{x'}$ and all other Gaussians in the catalogue. This also means that for our experiments we can pre-compute these elements without loss of generality (cf.~\cite{Schnitzer2012MutualProximity}).

\subsection{Hubness}
\label{subsec:Hubness}
The music recommender we look at in this work suffers from the consequences of hubness, leading to over a third of the songs in the catalogue never being recommended~\cite{Flexer2018MutualProximity}. So-called \emph{hubs} are often among the nearest neighbours of songs and are hence frequently recommended, leaving no room for \emph{anti-hubs} in nearest neighbour lists, and thus anti-hubs are never recommended. Hubs are usually determined by their \emph{k-occurrence} $O^k$, which is a measure describing the number of times a particular object (here: song) is within the $k$ nearest neighbours of all remaining objects that are part of a dataset. 
In what follows, we use a \emph{hub-size} of $5k$, i.e., for a song to be considered a hub it has to have a k-occurrence $O^k \geq 5k$. Note that the mean
$O^k$ across all objects is equal $k$, with $O^k$ significantly bigger than $k$ indicating a hub. Anti-hub songs never occur within the nearest neighbours of other songs, meaning they have a k-occurrence $O^k = 0$; all remaining songs, with $0 < O^k < 5k$, are \emph{normal} songs~\cite{Flexer2018MutualProximity}. Note once again that we let $k = 5$ in our experiments, resulting in a hub-size that equals $25$.

\subsection{Mutual Proximity}
\label{subsec:Mutual Proximity}
Mutual Proximity was previously proposed as a global scaling method improving the negative effect of hubness by repairing asymmetries of the neighbourhood relation between any two objects in a dataset~\cite{Schnitzer2012MutualProximity}. The main idea is to transform distances to a likelihood that one object $x$ is the nearest neighbour to another object $t$ (given the distribution of all distances to object $t$), and to combine it with the likelihood of also $t$ being the nearest neighbour of $x$. Only when both these likelihoods are high will \acs{mp} also achieve a high value. To estimate \acs{mp}, we follow the approach proposed by~\cite{Schnitzer2012MutualProximity} and use the empirical distribution to compute
\begin{equation}
\label{eq:mp}
    \operatorname{MP}(d_{x, t}) = \frac{\mid \{j: d_{x, j} > d_{x, t}\} \cap \{j: d_{t, j} > d_{t, x}\} \mid}{n}.
\end{equation}
Here $d_{x, t}$ is short for $D_{\mathrm{SKL}}(\mathcal{G}_{x}, \mathcal{G}_t)$ and corresponds to the symmetric \acs{kl} divergence between the Gaussians of $x$ and $t$; $n$ is the total number of objects. We turn \acs{mp} into a distance by defining $D_{\mathrm{MP}}(d_{x, t}) = 1 - \operatorname{MP}(d_{x, t})$.

\subsection{The Attack}
\label{subsec:The Attack}
In~\cite{Tismir2021}, an attack originally proposed by Carlini and Wagner~\cite{Carlini2018STT} for speech processing was used to compute adversarial examples for the real-world recommender system described above. The \ac*{cw} attack is a targeted white-box adversarial attack, meaning it requires full knowledge of a system under attack and aims at changing the output of the system to a predefined target $t$.
In our case, the objects to be adversarially modified are audio files representing Soundpark songs, and the \textit{adversarial perturbations} $\delta$ we wish to apply to these are modifications to the audio, in the form of additive noise, that are (hopefully) imperceptible. To achieve a desired distorted system prediction, which in our case is an increased number of times a particular song is recommended, we minimise a system loss with respect to the target output iteratively. In this modification of the attack, the target corresponds to a song within the dataset that is already a hub, and hence is often recommended. We simultaneously try to keep the added adversarial perturbation $\delta$ as small (or imperceptible) as possible by also minimising the squared L2-norm of a perturbation. 

In order to compute an adversarial perturbation $\delta$ for a particular song, with the aim of increasing its recommendations, we therefore try to minimise a combination of a system loss and the norm of the perturbation iteratively. This results in an optimisation objective and an update formula for $\delta$ as follows, which we can realise with gradient descent:
\begin{eqnarray}
\label{eq:cw_rec}
    L_\mathrm{total} &=& \|\delta_{ep}\|^2_2 + \alpha * D_{\mathrm{SKL}}(\mathcal{G}_{x + \delta_{ep}}, \mathcal{G}_t) \\
    \delta_{ep+1} &=& \clip_\epsilon (\delta_{ep} - \eta * \sign(\nabla_{\delta_{ep}} \ L_\mathrm{total})).
\end{eqnarray}

Note that a perturbation $\delta$ is here subscripted with the current epoch $ep$, and its updates are performed based on the $\sign$ of the gradient $\nabla_{\delta_{ep}}$ (w.r.t. $\delta_{ep}$) and the factor $\eta$. Updates are further clipped in each iteration to stay between $-\epsilon$ and $\epsilon$. The system loss of the music recommender here is represented by the \acs*{kl} divergence ($D_{\mathrm{SKL}}$), and $\mathcal{G}_x$ represents a Gaussian of a particular song $x$. The multiplicative factor $\alpha$ balances the focus of the optimisation between finding perturbations more quickly or less perceptible. For each $x$, the target song $t$ is chosen to be its closest hub song. A perturbation $\delta$ is updated until the attack is successful (stopping criterion), i.e., song $x$ has a $O^k \geq 25$, or until 500 update steps were made. The perturbation $\delta_0$ is initialised with zeros.

After a successful attack, the number of times a song is recommended by the system is therefore higher than before. This could be exploited in systems like the recommender of the FM4 Soundpark, as users can directly contribute to the song catalogue and could try to submit a perturbed version of their song in order to manipulate the recommender.

\subsubsection{Modified Mutual Proximity Attack}
\label{subsubsec:Modified Mutual Proximity Attack}
The attack described above uses knowledge about the song encoding and distance measure used by the recommender (via $\mathcal{G}$ and $D_{\mathrm{SKL}}$), but not about the specific defence (\acs*{mp}) it will face. In order to create a more difficult defence scenario for \acs*{mp}, we additionally incorporate knowledge of the defence into a modified adversarial attack.

More precisely, we attempt to attack the system by (I) leaving the main objective $L_\mathrm{total}$ unchanged (cf. Equation~(\ref{eq:cw_rec})) and only adapting the stopping criterion of the attack. The stopping criterion decides if an attack is successful, which is now only the case if the k-occurrence is at least 25 \emph{after} distances between all objects in a dataset are rescaled with \acs*{mp}. This is different to the original case, in which the attack was successful if the k-occurrence exceeded 25 \emph{without} applying \acs*{mp}. In what follows, we call this attack adaptation \emph{C\&W$^{mod}_{KL}$}. For the second adaptation (II), the stopping criterion is adapted in the same way; additionally, we change the objective $L_\mathrm{total}$ such that we minimise an approximation of the \acs*{mp} between a song $x$ and its target $t$ (cf. Equation~(\ref{eq:mp})) instead of the \acs*{kl} divergence. This approximation is necessary as \acs*{mp} itself is not differentiable. Including this knowledge results in an updated $L_\mathrm{total}$ in Equation (3) of
\begin{eqnarray}
\label{eq:adapted_cw}
    L'_\mathrm{total} &=& \|\delta_{ep}\|^2_2 + \alpha * \widetilde{D}_{\mathrm{MP}}(d_{x + \delta_{ep}, t}),
\end{eqnarray}
with
\begin{eqnarray}
\label{eq:mp_approx}
    \widetilde{D}_{\mathrm{MP}}(d_{x, t}) &=& 1 - \frac{\sum_i \bt_{i, t}(x) * \bt_{i, x}(t)}{n}, \\
    \bt_{i, t}(x) &=& \max(\tanh(d_{x, i} - d_{x,t}), 0).
\end{eqnarray}

Here $\widetilde{D}_{\mathrm{MP}}$ corresponds to an approximation of $D_{\mathrm{MP}}(d_{x, t})$, i.e., $\widetilde{D}_{\mathrm{MP}} \approx 1 - \operatorname{MP}(d_{x, t})$. The numerator in Equation (6) is the approximation of the numerator in Equation (\ref{eq:mp}), and consists of summation over $i$, where $i$ denotes all elements in our data catalogue. The constant $n$ in the denominator denotes the total number of objects in the catalogue. The function $\bt$ is a differentiable approximation of the \emph{bigger-than} function in Equation (\ref{eq:mp}); here the function $\tanh$ denotes the hyperbolic tangent function, and $\max$ returns the maximum of its two inputs. Subsequently, this modified attack will be denoted by \emph{C\&W$^{mod}_{MP}$}.


\section{Experiments}
\label{sec:Experiments}
To allow reproducibility of the experiments summarised in this work, the code as well as all necessary attack-parameters are available on Github\footnote{\url{https://github.com/CPJKU/hub_defence}}.

\subsection{Parameters of Attacks}
\label{subsec:Parameters of Attacks}
The adversarial attack proposed in~\cite{Tismir2021} and our modifications require a set of parameters, which influence the number of successful adversarial examples as well as their quality. More precisely, the parameters we need to choose are the clipping factor $\epsilon$, the multiplicative factor $\eta$ which determines the step-size of the gradient updates, and finally factor $\alpha$, which is responsible for controlling the focus of the attack on finding a higher number versus less perceptible adversaries. For the original attack, we tried to reproduce the results shown in ~\cite{Tismir2021} and used the proposed parameters, i.e. $\epsilon = 0.1, \eta = 0.001, \alpha = 25$.

For the remaining attacks, we performed a grid-search over various parameter combinations, and chose the settings in which we found the overall highest number of successful adversarial examples. For C\&W$^{mod}_{KL}$, this corresponds to $\epsilon = 1.0, \eta = 0.001, \alpha = 25$; For C\&W$^{mod}_{MP}$, we set $\epsilon = 1.0, \eta = 0.0005$, optionally with $\alpha = 100$. Note that this grid-search was done on the complete data base, since the white-box nature of the attacks requires full knowledge of the attacked system (here: all Gaussians). Additionally an attacker would naturally also have knowledge of the audio to be perturbed, hence a distinction into train and test data for parameter estimation, as is customary in most machine learning settings, is not necessary when evaluating such white-box attacks.

\subsection{Mutual Proximity as a Defence}
\label{subsec:Mutual Proximity as a Defence}
Before using \acs*{mp} as a defence, we investigate the vulnerability of the undefended recommender by applying the \acs*{cw} attack used in related work~\cite{Tismir2021}. The first line in Table~\ref{tab:mp_attack_res} shows the result of this attack. 

The columns in Table~\ref{tab:mp_attack_res} depict the nature of the attack, the number of initial hubs (i.e., before the attack), the number of adversarial hubs (i.e., songs for which the attack was successful), and the number of songs for which the attack was not successful (\# Non-hubs). The last two columns contain the average $\pm$ standard deviation of the \ac*{snr} and the k-occurrence of successful adversarial examples. We manage to successfully attack around $44.1\%$ of all files with an average \acs*{snr} of roughly 39.0dB, which is similar to the results shown in~\cite{Tismir2021}. Note that we use a subset of the data used in \cite{Tismir2021}.

\begin{table*}
    \centering
    \label{tab:mp_attack_res}
		\begin{tabular}{ l  c  c  c  c  c }
			\hline \rule{0pt}{10pt}
			Adaptation & \# Initial Hubs & \# Adversarial Hubs & \# Non-hubs & SNR & $O^k$ \\ [2pt]
            \hline \rule{0pt}{10pt}
            original & 202 (4.0\%) & 2,206 (44.1\%) & 2,592 (51.8\%) & 39.0 $\pm$ 5.1 & 41.7 $\pm$ 23.0 \\ \hline \rule{0pt}{10pt}
            C\&W$^{mod}_{KL}$ & 3 (0.1\%) &  126 (2.5\%) &  4,871 (97.4\%) & 28.3 $\pm$ 6.6 & 25.9 $\pm$ 1.3  \\
			\ C\&W$^{mod}_{MP}$ & 3 (0.1\%) &  18 (0.4\%) & 4,979 (99.6\%) & 42.9 $\pm$ 7.8 & 26.5 $\pm$ 1.5 \\
			\ C\&W$^{mod}_{MP}$ (no norm) & 3 (0.1\%) &  286 (5.7\%) & 4,711 (94.2\%) & 20.4 $\pm$ 9.2 & 26.1 $\pm$ 1.4 \\ [2pt]
            \hline
		\end{tabular}
	\caption{Results of the (adapted) \acs*{cw} attacks. The columns are the adaptation of the attack, the number of initial and adversarial hubs, the number of non-hubs after the attack, and the average \acs*{snr} and k-occurrence of successful adversaries.}
\end{table*}

Next, we use \acs*{mp} to defend against future adversarial attacks by integrating it in the music recommender. We first rescale the distances between all original (clean) objects in our dataset with \acs*{mp}. Beyond what is shown in Table~\ref{tab:mp_attack_res}, let us briefly look at some additional information displaying the impact of applying \acs*{mp} on the data. Before rescaling our dataset has 1,663 (33.3\%) anti-hub and 202 (4.0\%) hub songs, with a maximal k-occurrence of $O^k = 393$. This means that one third of the song collection is never recommended without \acs*{mp}. After rescaling however, this is reduced to 408 (8.2\%) anti-hubs and only 3 (0.1\%) hubs, with a maximal k-occurrence of $O^k = 35$. To further advocate the usage of \acs*{mp}, we can also look at the average retrieval accuracy $R^k$ as defined in~\cite{Schnitzer2012MutualProximity}, which increases from $43.7\%$ before \acs*{mp} to $47.7\%$ after \acs*{mp}.

After applying \acs*{mp}, we use the two adaptations of the attack in an attempt to find adversarial hubs for the defended system. Going back to Table~\ref{tab:mp_attack_res}, the second line shows that the success rate of the attack decreases from $44.1\%$ to only $2.5\%$ when we now try to minimise the \acs*{kl} divergence between a song and its target (C\&W$^{mod}_{KL}$). Also the \acs*{snr} of the successful adversaries decreases from on average $39.0$dB to $28.3$dB. If we use the second proposed adaptation (C\&W$^{mod}_{MP}$), and minimise the \acs*{mp} instead, the \acs*{snr} is higher ($42.9$dB), however we are only successful for a small percentage ($0.4\%$) of all files (see third line). In an attempt to find a larger number of successful adversaries to test the boundaries of \acs*{mp} as a defence, we repeat the attack using the adaptation C\&W$^{mod}_{MP}$ once more, yet leaving out the factor $\|\delta_{ep}\|^2_2$ restricting the norm of the perturbation (i.e., minimising $L'_\mathrm{total} = \widetilde{D}_{\mathrm{MP}}(d_{x + \delta_{ep}, t})$). The result is shown in the last line of Table~\ref{tab:mp_attack_res}. Disregarding the perceptibility of a perturbation during the optimisation process leads to a slightly higher success-rate of $5.7\%$, but also in a low average \acs*{snr} of 20.4dB. The perceptual differences of the different attacks can be examined in the supplementary material\footnote{\url{https://cpjku.github.io/hub_defence}}, where listening examples are provided. The song excerpts are chosen to represent good as well as bad examples for particular attacks in terms of their perceptibility. Note that perturbations with a \acs*{snr} of above $40dB$ are hardly perceptible; \acsp*{snr} between $40-20dB$ are perceptible, at least when compared to the original audio, but without noticeably changing the essence of the song. \acsp*{snr} lower than $20dB$ tend to become clearly perceptible or even disruptive.

\subsection{Post-Hoc Defence}
\label{subsec:Post-Hoc Defence}
Instead of integrating \acs*{mp} into the system directly, \acs*{mp} could also be applied post-hoc to rescale the distances between all files \emph{after} an attack , and hence post-process its recommendations. While in a real-world scenario it should be preferable to directly integrate the defence into a system to make an attack as difficult as possible, we want to briefly show that \acs*{mp} could also reduce the impact of an adversarial attack on an undefended system. For each of the $2,206$ hub songs we found when attacking the undefended system, we apply \acs*{mp} after one such song is added to the (otherwise clean) dataset, and observe how its k-occurrence and therefore hubness is changed due to the transformation. 

For $2,193$ out of the $2,206$ (i.e., $99.4\%$) files we manage to decrease the k-occurrence enough to revert them back to \emph{normal} songs (i.e., $0 < O^k < 25$). In other words, \acs*{mp} could also be a successful post-hoc defence for the \acs*{cw} attack on the music recommender.


\section{Discussion}
\label{sec:Discussion}
In this work we examined a specific real-world system to investigate the suitability of \acs*{mp} as an adversarial defence, which is why we briefly want to reason about how our findings could generalise to other settings. A limiting requirement of this defence is, due to its nature, that the attack is aimed at exploiting the \emph{hubness} issue in some way. However, we expect the proposed defence not to depend on the exact computation of adversarial examples, and hence to be useful against approaches different from the \acs*{cw}-like attack.

As previously mentioned, hubness has also been shown to be an issue in various different (recommender) systems, most importantly in systems based on collaborative filtering~\cite{Knees2014HubnessCollaborativeFiltering,hara2015reducing}. In other words, also systems like these are susceptible to attacks exploiting hubness, and could potentially be suited for and benefit from the proposed adversarial defence. Note here once more that hubness in a recommender system can lead to unfair recommendations, as a significant part of data is never recommended, while a relatively small part of the data is recommended very often (cf.~\cite{Flexer2018MutualProximity}); adversarial attacks that exploit hubness additionally amplify this effect on the fairness of recommendations, which is why a defence against them is crucial. However, as \acs*{mp} was shown to be able to reduce the hubness for various kinds of data (cf.~\cite{Feldbauer2019HubnessReductionComp}), we assume that the suitability of \acs*{mp} as a defence extends to a variety of different systems based on diverse data.

A further point for discussion is the potentially high computational complexity of the methods applied in this work. While the original definition of \acs*{mp} has quadratic complexity (w.r.t. the dataset size $n$), approximations that are linear in $n$ were previously proposed to allow an application of this method even to large datasets (cf.~\cite{Schnitzer2012MutualProximity,Feldbauer2018FastMP}). The high complexity also carries over to the modified adversarial attack, in which we therefore also used an approximation of \acs*{mp}. Nevertheless, the computational cost increases with growing dataset sizes, which is problematic in particular for the informed (modified) attack, as the \acs*{mp} needs to be computed in every iteration. The defence, however, requires additional computations of \acs*{mp} only if new data points are added to a catalogue.

Our work also connects to recent results establishing that, as the local intrinsic dimensionality of data increases, nearest neighbour classifiers become more vulnerable to adversaries \cite{amsaleg2017vulnerability}. Datasets are often embedded in spaces of higher dimensionality than is needed to capture all their information. The minimum number of features necessary to encode this information is called intrinsic dimensionality (see \cite{camastra2016intrinsic} for a recent review). It is well known that hubness also depends on a dataset’s intrinsic dimension~\cite{Radovanovic2010Hubs}. Future work should explore this possible new link of the concepts of adversaries, high dimensionality and hubness.


\section{Conclusion}
\label{sec:Conclusion}
In this work, we have evaluated an existing hubness reduction method (Mutual Proximity) as a defence against an adversarial attack on a real-world music recommender. Before the defence, the attack is able to artificially inflate playcounts of songs at the expense of other songs never being played, resulting in a very unfair recommender.  While the success rate of the attack is around $44.1\%$ before the defence with \acs*{mp}, we decrease it to $0.4\% - 5.7\%$ after the defence, despite incorporating knowledge of the defence in the attacks. In addition to making it more difficult to find successful adversarial perturbations, the defence also forces the attack to result in more perceptible perturbations. As we defend a specific real-world recommender in this work, we also discussed how this could generalise to other systems, and reason that it could be a suitable defence against diverse kinds of hubness-related attacks in the future.


\begin{acknowledgments}
This research was supported by the Austrian Science Fund (FWF, P 31988). GW's work is supported by the European Research Council (ERC) under the EU's Horizon 2020 research and innovation programme, grant agreement No \\ 101019375 (\textit{Whither Music?}). For the purpose of open access, the authors have applied a CC BY public copyright licence to any Author Accepted Manuscript version arising from this submission.
\end{acknowledgments} 

\bibliography{smc2022bib}

\end{document}